# On the limitations of optical characterization of intense ultrasound fields in the Raman-Nath regime

Gregory T Clement, Department of Physics, University of Rhode Island, Kingston

Email: gclement@bwh.harvard.edu

*The following technical report is excerpted from Appendix B of the author's PhD thesis, "Evaluation of transient and highly localized acoustic fields using acoustooptic phase contrast imaging" University of Rhode Island, ISBN 9780-599967922 (2000).*

Accurate interpretation of optical measurement of acoustic fields generally assumes two criteria are met regarding the acoustooptic interaction. First, the light-sound interaction must be weak enough to be considered in the Raman-Nath regime. That is, only diffracting orders up to the first order are considered. Second, the effects of optical refraction that take place over the width of the acoustic signal are generally treated as insignificant. The quantitative conditions dictating where each criterion is applicable can be expressed in terms of the Cook-Klein parameter and the Raman-Nath parameter. However, these conditions, which are the basis for most optical methods of acoustic field characterization, are incomplete. This technical report demonstrates a third requirement that is also necessary in order to assure the validity of the approach in high pressure characterization.

## INTRODUCTION

Accurate interpretation of phase-contrast images requires that two criteria are met regarding the acoustooptic interaction. First, the light-sound interaction must be weak enough to be considered in the Raman-Nath regime [1-5]. That is, only diffracting orders up to the first order are considered. Second, the effects of optical refraction that take place over the width of the acoustic signal must be insignificant. The quantitative conditions dictating where each criterion is applicable are generally expressed in terms of the Cook-Klein [6] parameter

$$Q = \frac{k_s^2}{k n_w} \tag{1}$$

and the Raman-Nath parameter

$$\nu = \frac{kCPL}{\beta}. \tag{2}$$

Here, the acoustic wavenumber is given by $k_s$, $L$ is the length of the interaction region along the optic axis, $k$ is the optic wavenumber, $n_\omega$ is the ambient optical index of refraction of the medium, $P$ is the acoustic pressure amplitude, $\beta$ is the adiabatic bulk modulus and $C$ is an elastooptic constant for the medium. If these conditions hold, a straight unrefracted ray approach (SURA) may be applied [7]. It has been argued in the literature [8] that the SURA approach is valid when (i) $Q \ll 1$ and (ii) $Q\nu \ll 2$ across the frequency spectrum of the acoustic signal. This technical report will demonstrate that a third requirement that, (iii) $\nu < 1$, is also necessary in order to assure the validity of the approach. To provide a concise quantitative description of these parameters, the elastooptic constant $C$ for water is derived from the Lorentz-Lorenz equation in Sec. II. This constant is then used in the mathematical justification for conditions (i) and (ii), here on in referred to as the first and second conditions respectively. These conditions are presented successively in Secs. III and IV. Finally, a discussion of the implications of the SURA approach in the optical transform plain is presented in Sec. V, where the necessity of condition (iii) is demonstrated. Implications on phase contrast filtering in this plane are also addressed in that section.

## II. ELASTOOPTIC CONSTANT

The relationship between acoustic pressure and the optical index of refraction may be derived from the Lorentz-Lorenz equation [9]. This equation describes the dependence of the index of refraction on the density of the medium and other material properties. The index is in general a tensor quantity. However for ideal fluids the equation becomes

$$\rho(n) = D\frac{n^2 - 1}{n^2 + 2}, \tag{3}$$

where $D$ is a constant independent of $n$. The derivative of Eq. (3) with respect to n gives,

$$d\rho = D\frac{6n}{(n^2 + 2)^2}dn. \tag{4}$$

Combination of Eq. (3) and Eq. (3) to eliminate $D$ yields

$$\frac{d\rho}{\rho} = \frac{6n}{(n^2 + 2)(n^2 - 1)}dn. \tag{5}$$

Over a region of slight variation, the right hand side of Eq. (5) is expressed in terms of the condensation $d\rho/\rho \to s$ while $dn \to \Delta n$. Assuming a linear relationship between small index changes and the condensation $\Delta n = Cs$, the elastooptic constant is obtained from Eq. (5) via the zeroth order expansion about the ambient index of refraction $n_0$,

$$C = \frac{(n_0^2 + 2)(n_0^2 - 1)}{6n_0}. \tag{6}$$

For water, $n_0 = 1.33$, giving a value of $C - 0.31$. Since the linearized acoustic pressure is $p = \beta s$, where $\beta$ is the adiabatic bulk modulus, using the combined assumptions of linear dependence of pressure and index change on condensation, the index may be expressed as

$$\Delta n = \frac{Cp}{\beta}. \tag{7}$$

with an adiabatic modulus water value of $\beta = 2.18 \times 10^9$ $Pa$.

### III. FIRST CONDITION

The condition that the Cook-Klein parameter be small, i.e. $Q \ll 1$, is an expression of the requirement that the sound field vary slowly relative to a wavelength of light and that the sound field must have a relatively thin waist. Justification of this requirement is seen by considering a planar light beam directed through a sound column such that the axes of propagation of the light is normal to the sound. Far from the source, a coherent light beam of width $w$ and wavelength $\lambda$ will spread over an angle $\lambda/w$ [10]. Singling out a small light ray bundle $\delta x$ from the beam, where the bundle is much smaller than the path length through the sound, $\delta x \ll L$, the bundle width across the sound then becomes

$$\delta x' = \delta x + L\frac{\lambda}{\delta x}. \tag{8}$$

If the width of the bundle across $L$ remains much smaller than the smallest appreciable wavelength of the sound spectrum, $\lambda_{Smin}$, refraction along the direction of acoustic propagation may be neglected. The bundle width in Eq. (8) minimizes to:

$$\delta x' = 2\sqrt{2L\lambda}.$$

Using this value, a slowly varying sound field requires that

$$2\sqrt{2L\lambda} \ll \lambda_{s_{min}}. \tag{9}$$

Rewriting this equation in terms of the wavenumber the condition becomes

$$\frac{4}{\pi}\frac{k_s^2}{k}L \ll 1. \tag{10}$$

The left hand side reduces to Eq (1), the Cook-KIein parameter, and since $4/\pi$ is on the order of one, and the condition may be written as:

$$Q \ll 1. \tag{11}$$

## IV. SECOND CONDITION

The condition that $Qv \ll 2$ is a quantitative expression of the requirement that the sound field must also be of sufficiently low intensity. The basis of the second condition is seen by again considering a ray bundle of width $\delta x$, traveling through a sound waveform. The condition will be determined by looking at a "worst case" situation, where the sound is assumed to be a rectangular column of width $L$ and constant pressure amplitude $P$ across the sound column. If the sound is propagating along the x-axis and the light along the z, the phase difference across the light bundle after traversing a distance $\Delta z$ through the sound will be

$$\Delta\phi(x) = k\Delta z \Delta n(x). \tag{12}$$

Since the bundle width is small, the gradient in the index of refraction across $\delta x$ is assumed constant. The additional path length taken by the faster moving edge of the bundle is then equal to $\Delta l = \Delta\phi/kn_0$, so that

$$kn_0\Delta l = k\Delta z\Delta n, \tag{13}$$

or with the limits $\lim_{\delta x \to 0}$ and $\lim_{\Delta z \to 0}$ the equation is rewritten,

$$\frac{\partial \alpha}{\partial z} = \frac{1}{n_0} \frac{\partial n}{\partial x} \tag{14}$$

where $\alpha = dx/dz$. Now by Eq. (7), a contribution to the index shift from a component of the acoustic signal of given wavenumber $k_s$ will be

$$n = \frac{CP_s}{\beta} e^{i(k_s x - \omega_s t)}. \tag{15}$$

Substitution of Eq. (15) into Eq. (14) then gives a constant as the second derivative with respect to x,

$$\frac{\partial^2 x}{\partial z^2} = \frac{CPk_s}{n_0 \beta}. \tag{16}$$

Here, $P$ denotes the complex wave of the single wavenumber component. The equation has a simple solution of

$$x = x_0 + bz + \frac{CPk_s}{2n_0 \beta} z^2. \tag{17}$$

Since x must be small relative to the sound wavelength dimensions,

$$\frac{CPk_s}{2n_0 \beta} L^2 << k_s^{-1}, \tag{18}$$

which is rewritten in terms of the Raman-Nath parameter and the Cook-Klein parameter

$$Q\nu << 2. \tag{19}$$

## V. ACOUSTOOPTICS IN THE OPTIC TRANSFORM PLANE: INTRODUCTION OF A THIRD CONDITION

Upon meeting each criterion described in the previous two sections, interpretation of the optic field in the Fourier plane is greatly simplified. This section examines the Fourier plane in this limit. It is demonstrated that the small-phase-shift assumption used to justify typically-encountered optical characterization of acoustic fields is justified; however, these conditions are may not be valid for high intensity fields. Specifically, it is shown that when Eqs. (11) and (19) hold *and* when a third condition on the Raman-Nath parameter, $\nu < 1$ for all wavevector components of the optical signal is introduced, diffraction orders above the first may be neglected.

The optical phase shift, immediately upon passing through the sound, may be decomposed into a Fourier cosine series of phase shifts,

$$\Delta\phi = \sum_{\gamma=-\infty}^{\infty} \Delta\phi_\gamma$$

due to each harmonic component of the acoustic field:

$$\Delta\phi_\gamma = \left[\int_0^L k_0 \Delta n(x) dz\right]_\gamma = k_0 n_0 L \cos(k_{s_\gamma} x - \omega_{s_\gamma} t + \theta_\gamma), \tag{20}$$

where $k_{s_\gamma}$ is the wavenumber of the component, $\omega_{s_\gamma}$ is its corresponding frequency, and $\theta_\gamma$ is its phase. The perturbed light in terms of its electric field is then

$$E_\gamma = E_0 e^{-ikn_0 L} e^{-ikn_0 L \cos(k_{s_\gamma} x - \omega_{s_\gamma} t + \theta_\gamma)}. \tag{21}$$

The light-sound interaction is traditionally viewed as an interaction which decomposes the light into a series of cylindrical waves [2]. This is accomplished quantitatively by applying the Jacobi-Anger expansion [11],

$$e^{iz\cos\theta} = \sum_{m=-\infty}^{\infty} i^m J_m(z) e^{im\theta}, \tag{22}$$

to Eq. (21). The field is then written as,

$$E_\gamma = E_0 e^{-ikn_0 L} \sum_{m=-\infty}^{\infty} (-i)^m J_m(\nu_\gamma) e^{im(k_{s_\gamma} x - \omega_{s_\gamma} t + \theta_\gamma)}, \tag{23}$$

where $\nu_\gamma$ is the Raman-Nath parameter $kL\Delta n_\gamma$. It is noted that a single harmonic component of the sound field results in an infinite set of orders of light, each propagating with a new wavevector, $k_m = K\hat{z} + mk\hat{x}$. Although an infinite number of orders is produced, it is important to note that the relative intensity of each order in Eq. (23) is simply

$$I_m = \left[\frac{J_m(\nu_\gamma)}{J_0(\nu_\gamma)}\right]^2. \tag{24}$$

In the special case $\nu_\gamma \ll 1$, orders of Bessel functions above the first may be neglected. This observation represents the third condition proposed here. For example, for a parameter $\nu_\gamma = 0.1$, neglecting higher orders would introduce only 1% error, whereas intensities cannot be readily approximated for parameters above 1.

Based on this general property of Bessel function about zero [12], and assuming $\nu_\gamma \ll 1$ holds for all components, Eq. (23) reduces to

$$E_\gamma = E_0 e^{-ikn_oL}(1 + iJ_1(\nu_\gamma)e^{ik_{s\gamma}x - \omega_{s\gamma}t + \theta_\gamma}). \tag{25}$$

When Eq. (25) is then expanded in a McLaurean series to the second order for $\nu_\gamma$, the following result is obtained:

$$E_\gamma = E_0 e^{-ikn_oL}\left(1 + \frac{iLk\Delta n_\gamma}{2}e^{ik_{s\gamma}x - \omega_{s\gamma}t + \theta_\gamma}\right). \tag{26}$$

This is the exact solution of Eq. (5) in [13] for the harmonic case. The solution readily generalizes to the case on an arbitrary acoustic waveform,

$$E = E_0 e^{-ikn_oL}(1 + i\Delta\phi), \tag{27}$$

where

$$\Delta\phi = \sum_{\gamma=-\infty}^{\infty} \frac{iLk\Delta n_\gamma}{2}e^{ik_{s\gamma}x - \omega_{s\gamma}t + \theta_\gamma}.$$

**REFERENCES**

[1] C.V Raman and N.S. Nath, "The Diffraction of light by Sound Waves of High Frequency: Part I," *Proc. Ind. Acad. Sci. (A)* 2, 406-412 (1935).

[2] C.V Raman and N.S. Nath, "The Diffraction of light by Sound Waves of High Frequency: Part II," Proc. Ind. Acad. Sci (A) 2, 413-420 (1935).

[3] C.V Raman and N.S. Nath, "The Diffraction of light by Sound Waves of High Frequency: Part III," *Proc. Ind. A cad Sci. (A)* 3, 75-84 (1936).

[4] C.V Raman and N.S. Nath, "The Diffraction of light by Sound Wives of High Frequency: Part IV, " *Proc. Ind. Acad. Sci. (A)* 3, 119-125 (1935).

[5] C.V Raman and N.S. Nath, "The Diffraction of light by Sound Waves of High Frequency: Part V" Proc. Ind. Acad. Sci. (A) 3,459-465 (1935).

[6] WR. Klein and B.D. Cook,"Unified Approach to Light Diffraction," *IEEE Trans, on Son. and Ultra.,* 14(3), 123-133 (1967).

[7] Adrian Korpel, *Acousto-optics,* M. Dekker, New *Yotk* (1988).

[8] Adrian Korpel, "Eikonal Theory of Bragg Diffraction imaging," *Acoustical Holography* 2, Plenum Press, New York (1970).

[9] Born and E. Wolf, *Principles of Optics,* Wiley, New York, (1975).

[10] Robert Adler, "Interaction between light and sound," IEEE Spectrum 4, 42-54 (1967).

[11] Sadri Hassani, *Foundations of mathematical physics* , Ailyn and Bacon, Boston (1991).

[12] Milton Abramowitz and Irere Stegun, Editors, *Handbook of Mathematical Functions,* Dover, New York (1965).

[13] G.T. Clement, and S.V. Letcher "Optical imaging of transient acoustic fields using a phase contrast method" arXiv:1407.4758 [physics.optics]